\documentstyle[amssymb,epsf,pstricks]{article}
\title{The Role of Innovation within Economics}
\author{Russell K. Standish\\High Performance Computing Support Unit\\The
University of New South Wales,\\Sydney, 2052}

\newcommand{\reals}{{\Bbb R}}

\newcommand{\br}{\mbox{\boldmath{$r$}}}          %reproductive rate
\newcommand{\bbeta}{\mbox{\boldmath{$\beta$}}}   %interaction matrix
\newcommand{\bmu}{\mbox{\boldmath{$\mu$}}}       %mutation rate
\newcommand{\bn}{\mbox{\boldmath{$n$}}}          %species density
\newcommand{\nsp}{\mbox{$n_{\rm sp}$}}           %no. of species           

\begin{document}
\maketitle

\section{Introduction}

Comparative statics is a special case of dynamics, in which a unique
stable equilibrium is assumed to exist, and two equilibria ---
determined by different parameter values --- are compared. Similarly,
dynamics is a special case of an evolutionary process, in which the
degrees of freedom of the system are held constant. Marshall, whose
{\em Principles}\cite{Marshall20} played a large r\^ole in setting
economics upon the comparative statics course, was nonetheless aware of
the primacy of the evolutionary paradigm when he wrote on page xiv
that:

\begin{quote}
The Mecca of the economist lies in economic biology rather than in
economic dynamics. But biological conceptions are more complex than
those of dynamics; a volume on Foundations must therefore give a
relatively large place to mechanical analogies; and frequent use of
the term `equilibrium', which suggest something of a statical
analogy.\ldots 
The modern mathematician is familiar with the notion that dynamics
includes statics. If he can solve a problem dynamically, he seldom
cares to solve it statically also\ldots But the statical solution has
claims of its own. It is simpler than the dynamical; it may afford
useful preparation and training for the more difficult dynamical
solution; and it may be the first step towards a provisional and
partial solution in problems so complex that a complete dynamical
solution is beyond our attainment.
\end{quote}

%At the time Marshall penned these words, Darwin's theory of evolution
%had captured the scientific and public imagination. It defined state
%of the art biology at the time. The theory had enormous explanatory
%power, but little predictive power. By contrast, the dynamics of the
%time (nowadays called {\em classical dynamics}) had been refined over
%the previous two centuries, culminating in the likes of Laplace and
%Poincar\'e. This theory had enormous predictive value, so much so
%that the prevalent view was that if the initial states were known
%precisely, the behaviour of the system can be predetermined for all
%time.

One century later, intellectual progress has enabled us to contemplate
the construction of evolutionary models of processes which could once
only be modeled statically.

%Faced with the problem of understanding the world of human affairs,
%and having two such starkly different world views available, it is not
%surprising that Marshall would claim that economics is better
%understood as a biological (meaning evolutionary) system. Now, a
%century later, a bridge is being built between these two world
%views. This leads to the possibility of a predictive economics that is
%cognisant of its evolutionary heredity. Progress in the understanding
%of ecology illustrates this point.

%Until the 1970s, ecology was studied statically. 
Early ecological theories followed a similar path of development.
Interactions between organisms were studied in order to determine the
population densities that kept the ecology in equilibrium and the only
form of dynamics considered was instantaneous movement from one
equilibrium to another. It was always assumed that an ecology was
always near a stable equilibrium, in spite of counter examples such as
the stable limit cycles in the Lotka-Volterra predator-prey
system\cite{Maynard-Smith74}.

In the 1960s, interest in dynamical systems was reawakened, and the
study of {\em Chaos} was born. These techniques were applied to
ecology in the 1970s, and it was realised that the situation was
vastly more complicated, with not only limit cycles being possible,
but also entities called {\em strange attractors}. Over the years,
appreciation has grown that the stable equilibrium is more of an
exception than the norm.

In the mid 1980s, Raup and Sepkowski\cite{Raup-Sepkowski86} noticed
certain statistical regularities with the pattern of speciation and
extinction in the fossil record. These were later shown to be in the
form of a power law\cite{Sole-etal97,Newman96}, i.e the pattern
of extinctions was such that the frequency of an extinction event of
size $x$ is proportional to $x^{-\alpha}$, where $\alpha$ is a small
positive constant. Power laws crop up in many different
areas.\cite{Schroeder91}

Per Bak\cite{Bak-etal88} studied a model of a sandpile that had a
continuous stream of sand added to the top.  Bak's process goes by the
name of {\em self-organised criticality}, and he, along with
Kauffman\cite{Kauffman93} promote this as an explanation for Raup's
data. It now seems likely that evolution is an endogenously
self-organised critical process\footnote{Alternatively, perhaps
  exogenous influences (such as volcanos, meteorites, climatic
  fluctuations, supernovae etc) obey a power law spectrum. It is quite
  possible that both endogenous and exogenous effects contribute to
  the statistical properties of biological evolution and that a future
  research program will look at untangling the two effects. To this
  effect, Newman\cite{Newman97b} provides an interesting perspective.}

In my work, which I will introduce later, I have demonstrated the
existence of self-organised criticality in a model evolutionary
ecology. The criticality is quite a robust feature over a wide range
of input parameters. This indicates that that evolutionary systems are
typically endogenously critical.

In summary, we have a hierarchy of approaches, from the static, to the
dynamic to the evolutionary:
\begin{displaymath}
\mbox{\rm statics} \subset \mbox{\rm dynamics} \subset \mbox{\rm evolution}
\end{displaymath}
The state of ecological thought has followed this chain from left to
right, as computational techniques have improved to embrace the
computationally more difficult dynamical models, and then the even
more computationally difficult evolutionary models. It is to be hoped
that the same path will be followed by economic modeling.

%It is perhaps a truism that economics and
%ecology must have something in common. After all, ecology is all about
%the distribution of resources --- energy, material, territory etc,
%just as economics  deals with much the same
%problems in the human sphere, albeit with the added complications of
%money and prices.
%
%In biology, the fundamental theoretical plank is the theory of
%evolution\footnote{Evolution of species is a fact, as testified by the
%  fossil record. However, there are also theories of evolution, which
%  attempt to explain that fact.}, as initiated by Darwin, and
%developed considerably throughout this century. Evolutionary theory is
%built on the twin strands of mutation and ecology. Mutation provides
%the ``natural variation'' on which ``natural selection'', the outcome
%of ecological interactions, acts. Just as ecology is a central
%component of biological evolution, so is economics an equally
%important component of technological and economic evolution.

There is a clear parallel between the development of ecological
thought and that of economic thought. Dynamics was introduced to
economics and championed by Kaldor, Goodwin et al. in the 1950s and
60s; and boosted by the developments in nonlinear analysis in the
1980s, through the work of people like Blatt\cite{Blatt83}. This
volume testifies to a growing industry of dynamical economic modeling.
However, perhaps now is the time to embark on the next rung up the
ladder, and approach economics from an evolutionary point of view.
Then perhaps we might be catching a glimpse of the Mecca that Marshall
referred to a hundred years ago.

This paper first considers a general dynamical system undergoing
evolution of the determining equation, then outlines {\em Ecolab}, a
model of species interaction undergoing evolution, finally introducing
a possible economics model based on the von Neumann model with
evolution of the interaction coefficients.

\section{Linearisation of a Dynamical System}

Our launchpad for a theory of evolutionary systems is  dynamical systems
theory\cite{Hirsch-Smale74}. Typically this
will be manifested in a first order nonlinear differential equation of the form
\begin{displaymath}
\dot{\bf x}={\bf f}({\bf x})
\end{displaymath}
where ${\bf x}\in\reals^n$ and ${\bf f}:\reals\longrightarrow\reals$.
The dot refers to the derivative with respect to time.

Dynamical systems theory starts by considering the equilibria  of the
system, i.e. the points $\hat{\bf x}$ such that ${\bf f}(\hat{\bf x})=0$.
Then in the neighbourhood of $\hat{\bf x}$, the behaviour of the system
is determined by the linear approximation
\begin{displaymath}
\dot{\bf x}=D{\bf f}|_{\hat{\bf x}}\cdot({\bf x}-\hat{\bf x}).
\end{displaymath}
The stability of $\hat{\bf x}$ is determined by the negative
definiteness of $D{\bf f}|_{\hat{\bf x}}$\footnote{$D{\bf
f}|_{\hat{\bf x}}$ is negative definite if ${\bf x}\cdot D{\bf
f}|_{\hat{\bf x}}\cdot{\bf x}<0$ for all ${\bf x}$. This also implies
that all eigenvalues of $D{\bf f}$ have negative real part.}.  This
condition imposes $n$ inequalities on the system constraining the form
of ${\bf f}$. There may additionally be a further $n$ inequalities for
$\hat{\bf x}$ to be a meaningful solution, for example if the
components of $\hat{\bf x}$ are production values, every component of
$\hat{\bf x}$ must be non-negative.

General equilibrium economics has attempted to find the conditions
under which a unique, stable equilibrium will exist. This needn't be
the case, and interesting (i.e. bounded) behaviour can take place
around unstable equilibria, in the form of limit cycles or even the
strange attractors beloved of chaos theorists. A favourite model of
the latter researchers is the logistic equation, which first arose in
a biological context\cite{May76b}, but has been applied to
economics\cite{Thompson92} amongst other things.

That limit cycles and chaotic behaviour can be observed in economics
is a view that should have by now been accepted\cite{Blatt83}.
However, the question remains as to whether this behaviour is
pathological, i.e. whether linear neoclassical theory is applicable
in most cases, and the only remaining difficulty is determining if
linear theory applies to a specific economy (the ``econometric
problem''), or whether chaotic behaviour is indeed the norm.

\section{Limits of Linear Economic Theory}

Returning to the question of stability, the fact that $2n$
inequalities must be satisfied would imply that a randomly chosen
$n$-dimensional economics model would have a probability of $4^{-n}$ of
a given equilibrium being stable. The situation does look bleak, but
economics are not generated randomly in the real world, rather they
are the result of an evolutionary process. We need to examine the
process of cultural evolution to answer this question.

The analogue of mutation in biology would be innovation in economics,
as a new process or technique introduced into production, or as a new
form of marketing, or a new company with a somewhat unusual approach
to doing business. The effect of innovation is to add new degrees of
freedom to the dynamical system, which usually will destabilise the
system. The system will then tend to evolve so as to lose some of the
degrees of freedom, by for example old production techniques being
abandoned, or companies going bankrupt. This is analogous to species
becoming extinct in the natural world.

Lets consider what happens to the largest eigenvalue of $D{\bf
  f}|_{\hat{\bf x}}$.  Suppose initially, the system has a stable
equilibrium, in which case all the eigenvalues have negative real
part. As innovations are added to the system, the largest eigenvalue
will increase towards zero. As it passes zero, the system
destabilises, and the system will start to exhibit limit cycles or
chaotic behaviour. As further innovations are added to the system, a
property called {\em permanence}\footnote{that there is a set of
  points ${\bf x}_0$ whose trajectories ${\bf x}(t)$ always remain
  away from the boundary, i.e. $x_i(t)>\delta\;\;\exists\delta>0$} is
no longer satisfied, and some event such as a bankruptcy will occur to
remove active processes from the system. This will restore permanency
to the system, and possibly even stability. Such a process is called
{\em self-organised criticality}\cite{Bak-etal88} which gives rise to
a power law spectrum of the booms and busts, successful innovations
and bankruptcies.\footnote{This really implies that conclusion derived
  from comparative static economic analysis are almost never valid,
  except perhaps on sufficiently small time scales while the maximum
  eigenvalue of $D{\bf f}|_{\hat{\bf x}}$ is negative.}

\section{Ecolab and the Dynamics of Evolution}

This section outlines a model of an evolving
ecology\cite{Standish94,Standish96} that is analogous to an economic
system with input-output relations of production and product
innovation.  The ecology is described by a generalised Lotka-Volterra
equation, which is perhaps the simplest ecological model to use.
\begin{equation} \label{lotka-volterra}
\dot{n}_i = r_i n_i + \sum_{j=1}^{\nsp}\beta_{ij}n_in_j
\end{equation}
Here \br{} is the difference between the birth rate and death rate for
each species, in the absence of competition or symbiosis. \bbeta{} is
the interaction term between species, with the diagonal terms
referring to the species' self limitation, which is related in a
simple way to the carrying capacity $K_i$ for that species in the
environment by $K_i=-r_i\beta_{ii}$.  In the literature (eg
Strobeck\cite{Strobeck73}, Case\cite{Case91}) the interaction terms
are expressed in a normalised form, $\alpha_{ij}=-K_i/r_i\beta_{ij}$, and
$\alpha_{ii}=1$ by definition. \bn{} is the species density.

These equations are simulated on a simulator called {\em
  Ecolab}.\cite{Ecolab-Tech-Report} The vectors \bn{} and \br{} are
stored as dynamic arrays, the size of which (i.e. the system
dimension) can change in time.  The interaction array is stored in
row/column sparse form, consisting of the four arrays {\tt diag},
{\tt val}, {\tt row} and {\tt col}. Equation (\ref{lotka-volterra})
can be written as:
\begin{verbatim}
      tmp[row] = beta.val * n[beta.col];
      n += (r + beta.diag + tmp) * n;
\end{verbatim}

This code makes up the {\tt generate} operator in the Ecolab
system. Other operators include {\tt compact}, which removes species
that have become extinct from the system (to optimise computational
performance) and {\tt mutate}, which adds a certain number of new
species to the system, according to a specific algorithm to be
discussed later. The operators can be called from a scripting language
called TCL\cite{Ousterhout94}, that allows different types of
experiments to be performed without recompiling the code.

Before discussing the mutation algorithm in more detail, equation
(\ref{lotka-volterra}) must be analysed to determine the conditions
\bbeta{} must satisfy for the system to be real, and also to determine
the different regimes of dynamics, from the linear (stable
equilibrium) case, to limit cycles and chaos to the actual breakdown
of the ecosystem.

\subsection{Linear Analysis}

Linear analysis starts with the fixed point of equation (\ref{lotka-volterra})
\begin{equation}\label{fixed point}
\hat{\bn} = -\bbeta^{-1}\br,
\end{equation}
where $\dot{\bn}=0$. There is precisely one fixed point in the
interior of the space of population densities (i.e. \bn{} such that
$n_i>0$) provided that all components of $\hat{\bn}$ are
positive, giving rise to the following inequalities:
\begin{equation}\label{positive species}
\hat n_i = \left(\bbeta^{-1}\br\right)_i>0,\;\; \forall i
\end{equation}
This interior space is denoted $\reals_+^{\nsp}$ mathematically.

There may also be fixed points on the boundary of $\reals_+^{\nsp}$,
where one or more components of \bn{} are zero (corresponding to an
extinct species). This is because the subecology with the living
species only (i.e. with the extinct species removed) is equivalent to
the full system.

The {\em stability} of this point is related to the
negative definiteness of derivative of $\dot{\bn}$ at $\hat{\bn}$. The
components of the derivative are given by
\begin{equation}\label{derivative}
\frac{\partial\dot{n}_i}{\partial n_j} =
\delta_{ij}\left(r_i+\sum_k\beta_{ik}n_k\right) + \beta_{ij}n_i
\end{equation}
Substituting eq (\ref{fixed point}) gives
\begin{equation}
\left.\frac{\partial\dot{n}_i}{\partial n_j}\right|_{\hat{\bn}}=
-\beta_{ij}\left(\bbeta^{-1}\br\right)_i
\end{equation}

Stability of the fixed point requires that this matrix should be
negative definite. Since the $\left(\bbeta^{-1}\br\right)_i$ are
all negative by virtue of (\ref{positive species}), this is equivalent
to \bbeta\ being negative definite, or equivalently, that its \nsp\
eigenvalues all have negative real part. Taken together with the
inequalities (\ref{positive species}), this implies that $2\nsp$
inequalities must be satisfied for the fixed point to be stable. This
point was made by Strobeck\cite{Strobeck73}, in a slightly different
form. (Note that Strobeck implicitly assumes that
$\sum_ir_i\hat{n}_i / K_i >0$, so comes to the conclusion that
$2\nsp-1$ conditions are required.)  If one were to randomly pick
coefficients for a Lotka-Volterra system, then it has a probability of
$4^{-\nsp}$ of being stable, i.e. one expects ecosystems to become
more unstable as the number of species increases\cite{May74}.

\subsection{Permanence}

Whilst stability is a nice mathematical property, it has rather less
relevance when it comes to real ecologies. For example the traditional
predator-prey system studied by Lotka and Volterra has a limit
cycle. The fixed point is decidedly unstable, yet the ecology is {\em
permanent} in the sense that both species' densities are larger than
some threshold value for all time. Hofbauer et
al. \cite{Hofbauer-etal87} and Law and Blackford\cite{Law-Blackford92}
discuss the concept of {\em permanence} in Lotka-Volterra systems,
which is the property that there is a compact absorbing set ${\cal
M}\subset\reals^{\nsp}_+$ {\em i.e} once a trajectory of
the system has entered ${\cal M}$, it remains in ${\cal M}$. They derive a sufficient
condition for permanence due to Jansen\cite{Jansen87} of the form:
\begin{equation}\label{Jansen}
\sum_ip_if_i(\hat{\bn}_B) =
\sum_ip_i(r_i-\sum_j\beta_{ij}\hat{n}_{Bj}) > 0, \;\; \exists p_i>0 
\end{equation}
for every $\hat{\bn}_B$ equilibrium points lying on the boundary
($\hat{n}_{Bi}=0 \;\; \exists i$), provided the system is {\em
  bounded} (or equivalently {\em dissipative}). This condition is more
general than stability of the equilibrium --- the latter condition
implies that a local neighbourhood of the equilibrium is an absorbing
set. Also, the averaging property of Lotka-Volterra systems implies
that the equilibrium must lie in the positive cone $\reals^{\nsp}_+$.
So (\ref{positive species}) must still hold for permanence.

Consider the boundary points $\hat{\bn}_B$ that are
missing a single species $i$. Then Jansen's condition for these
boundary points is
\begin{equation}\label{single-deficiency}
r_i-\sum_j\beta_{ij}\hat{n}_{Bj}>0.
\end{equation}
This set of conditions is linearly independent. Let the number of such
boundary points be denoted by $n_B\leq\nsp$. Then the set of
conditions (\ref{Jansen}) will have rank $n_B\leq\nu\leq\nsp$ (the
number of linearly independent conditions), so the system has at
most probability $2^{-\nsp-\nu}$ of satisfy Jansen's  permanence condition
if the coefficients are chosen uniformly at random. As stability is
also sufficient for permanence, the probability lies between
$4^{-\nsp}$ and $2^{-\nsp-\nu}$.

Another rather important property is {\em resistance to
invasion}.\cite{Case91} Consider a boundary equilibrium
$\hat{\bn}_B$. If it is proof against invasion from the missing
species, then the full system cannot be permanent. For the boundary
points that miss a single species, this implies that condition
(\ref{single-deficiency}) is necessarily satisfied for permanence,
along with  (\ref{positive species}). The probability of
permanence is then bounded above by $2^{-\nsp-n_B}$.

Thus whilst a randomly selected ecology is more likely to be permanent
than to have a stable equilibrium, the likelihood decreases
exponentially with increase in species number.

\subsection{Boundedness}

It is necessary that the ecology be {\em bounded}, ie that $\sum n_i <
N \;\;\exists N, \;\; \forall t>0$. This requires 
\begin{equation}\label{boundedness}
\sum_i\dot{n_i}=\br\cdot\bn + \bn\cdot\bbeta\bn < 0, \;\;\forall \bn:
\sum_in_i>N \;\;\exists N
\end{equation}

As \bn\ becomes large in any direction, this functional is dominated
by the quadratic term, so this implies that 
\begin{equation}\label{boundedness3}
\bn\cdot\bbeta\bn\leq0 \;\;\forall\bn: n_i>0.
\end{equation}
If strict equality holds, then $\br\cdot\bn<0$.
Negative definiteness of \bbeta\ is sufficient, but not necessary for
this condition. Another sufficient condition is to require
$\forall i,j, \beta_{ii}<0$ and  $\beta_{ij}+\beta_{ji}\leq0$, which is used in the
current study. This condition is satisfied by the Predator-Prey
equations, and so does allow multi-trophic systems to be built, but
does not allow the possibility of symbiosis. Its main advantage is its
simplicity of implementation, along with the range of interesting
(i.e. non limit point) behaviour it encompasses.

\subsection{Mutation}

Adding mutation involves adding an additional operator to
equation (\ref{lotka-volterra}) 
\begin{equation}
\dot{\bn} = \br*\bn + \bn*\bbeta\bn + {\tt mutate}(\bmu,\br,\bn)
\end{equation}
where $*$ refers to elementwise multiplication.

The mutation operator must generate new degrees of freedom $i>\nsp$
(where \nsp\ is the number species currently in the ecology), somehow
defining the new ecological coefficients $\{r_i| i>\nsp\},
\{\beta_{ij}|i>\nsp\; \mbox{or}\; j>\nsp\}$ from the previous state of
the system. In reality, there is another layer (hidden in equation
(\ref{lotka-volterra}) called the genotypic layer, where each organism
has a definite genotype. There is a specific map from the genotypic
layer to the space of ecological coefficients (hereafter called the
phenotypic layer) called the {\em embryology}. Then the mutation
operator is a convolution of the genetic algorithm operations
operating at the genotypic layer, with the embryology.

A few studies, including Ray's Tierra world, do this with an explicit
mapping from the genotype to to some particular organism property
(e.g. interpreted as machine language instructions, or as weight in a
neural net). These organisms then interact with one another to
determine the population dynamics. In this model, however, we are
doing away with the organismal layer, and so an explicit embryology is
impossible. The only possibility left is to use a statistical model of
embryology. The mapping between genotype space and the population
parameters \br, \bbeta{} is expected to look like a rugged landscape,
however, if two genotypes are close together (in a Hamming sense) then
one might expect that the phenotypes are likely to be similar, as
would the population parameters. This I call {\em random embryology
  with locality}. Here, we tend to idealise genotypes as bit strings,
although strings over an arbitrary alphabet (eg the four DNA bases
ACGT) can equally be considered. \footnote{The Hamming distance is the
  number of bits (bases) that differ between the two strings. So for
  example if a single bit has been removed from one string, the
  Hamming distance is one.}

  In the simple case of point mutations, the probability $P(x)$ of any
child lying distance $x$ in genotype space from its parent follows a
Poisson distribution, as this is the distribution of the number of bit
flips, or deletions that might occur with a point mutation. Random
embryology with locality implies that the phenotypic parameters are
distributed randomly about the parent species, with a standard
deviation that depends monotonically on the genotypic
displacement. The simplest such model is to distribute the phenotypic
parameters in a Gaussian fashion about the parent's values, with
standard deviation proportional to the genotypic displacement. This
constant of proportionality can be conflated with the species'
intrinsic mutation rate, to give rise another phenotypic parameter
\bmu.  It is assumed that the probability of a mutation generating a
previously existing species is negligible, and can be ignored. We also
need another arbitrary parameter $\rho$, ``species radius'', which can
be understood as the minimum genotypic distance separating species,
conflated with the same constant of proportionality as \bmu.

\begin{figure}
\begin{pspicture}(0,0)(15,8)
\rput(6,4){\epsfbox{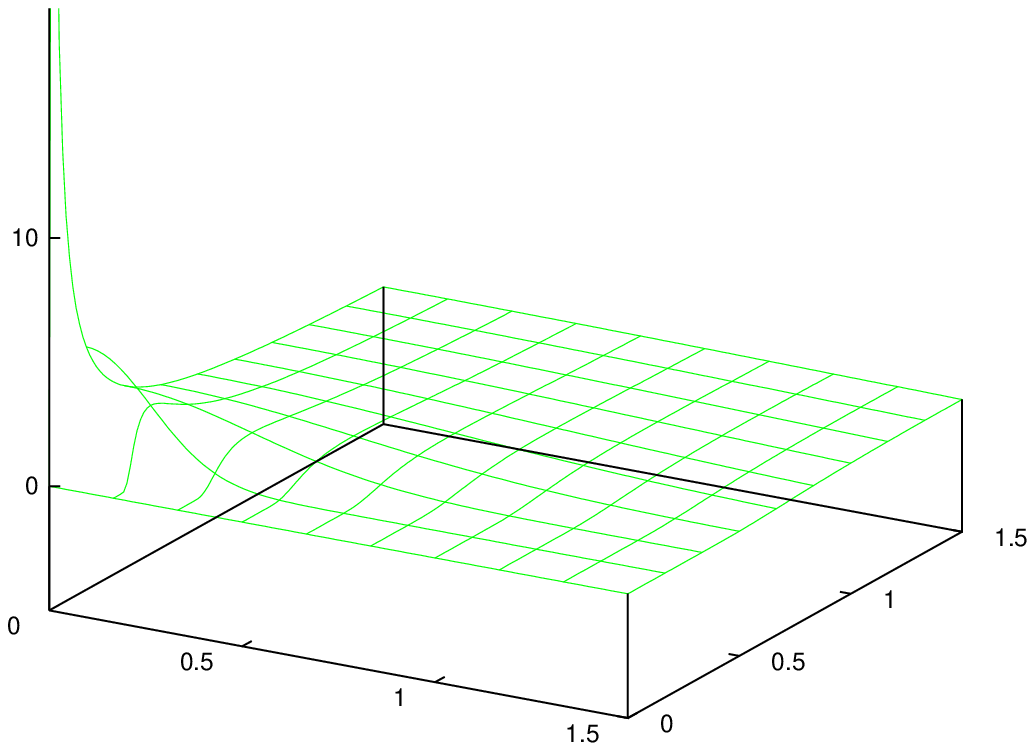}}
\psline{->}(.4,1.2)(6.2,0)
\psline{->}(8,0.4)(11.2,2.4)
\rput{-8.1}(3,.4){$p$}
\rput{30}(10,1){$g/\mu$}
\end{pspicture}
\caption{Probability distribution of the relation between genotype
difference and the corresponding phenotype difference}
\label{eco-embryo}
\end{figure}

We may represent the Ecolab embryology as a probability distribution
$f(p,g)=\sqrt{\frac2\pi}\frac{\mu e^{-\left(\frac{\mu
p}{2g}\right)^2}}{g}$, where $p=|r_i-r_j|/|r_i|$ or
$p=|\beta_{ik}-\beta_{jk}|/|\beta_{ik}|$ is the distance between two
species' phenotypic parameters, and $g$ is the difference between the two
genotypes. Figure \ref{eco-embryo} shows the general form of this
probability distribution.

\begin{figure}
\begin{pspicture}(0,0)(15,10)
\rput(6,5){\epsfbox{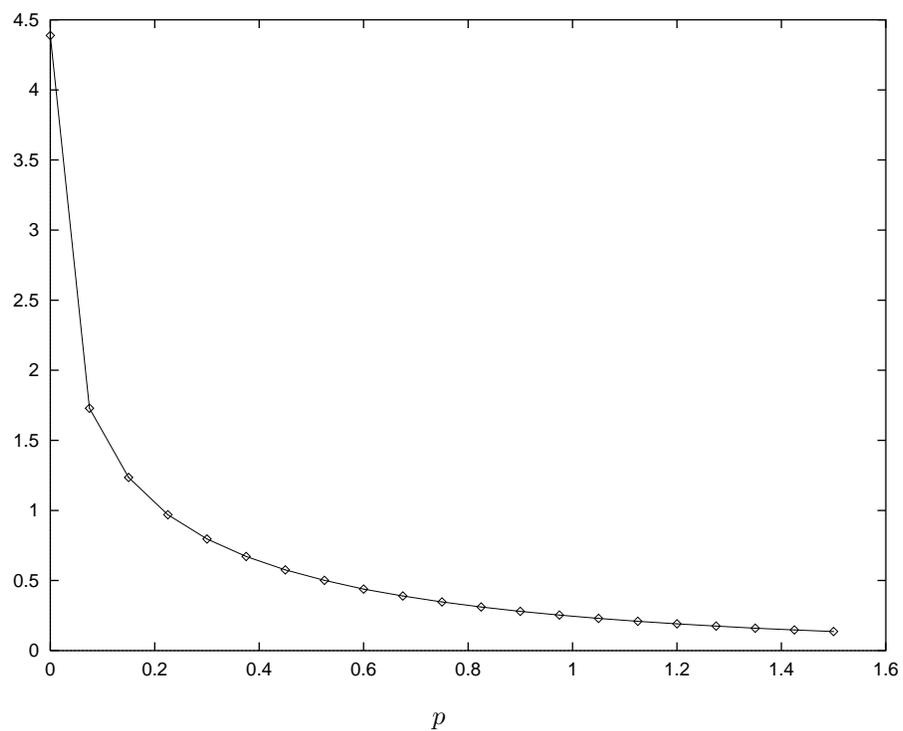}}
\rput(6,0){$p$}
\end{pspicture}
\caption{The probability distribution of a
mutant phenotypical coefficient about that of its parent's value. This
has been calculated by numerical integration from formula
(\ref{mut-eco-embryo-formula}). Note that the curve actually diverges
at 0.}
\label{mut-eco-embryo}
\end{figure}

Figure \ref{mut-eco-embryo} shows the probability distribution of a
mutant phenotypical coefficient about that of its parent's value. This
is given by 
\begin{equation}\label{mut-eco-embryo-formula}
\int_0^\infty \sqrt{\frac2\pi}\frac{e^{-g/\mu-\left(\frac{\mu
p}{2g}\right)^2}\mu}{g}dg.
\end{equation}

In summary, the mutation algorithm is as follows:
\begin{enumerate}
\item The number of mutant species arising from species $i$ within a
timestep is $\mu_i r_i n_i/\rho$. This number is rounded
stochastically to the nearest integer, e.g. 0.25 is rounded up to 1
25\% of the time and down to 0 75\% of the time.

\item Roll a random number from a Poisson distribution
$e^{-x/\mu+\rho}$ to determine the standard deviation $\sigma$ of phenotypic
variation. 

\item Vary \br\ according to a Gaussian distribution about the
parents' values, with $\sigma r_0$ as the standard deviation,
where $r_0$ is the range of values that \br\ is initialised
to, ie $r_0 = \left.\max_ir_i\right|_{t=0} -
\left.\min_ir_i\right|_{t=0}$. 

\item The diagonal part of \bbeta\ must be negative, so vary \bbeta\
according to a log-normal distribution. This means that if the old
value is $\beta$, the new value becomes
$\beta'=-\exp(-\ln(\beta)+\sigma)$. These values cannot 
arbitrarily approach 0, however, as this would imply that some species make
arbitrarily small demands on the environment, and will become infinite
in number. In ecolab, the diagonal interactions terms prevented from
becoming larger than $-r/(.1*{\tt INT\_MAX})$, where $r$ is the
corresponding growth rate for the new species.

\item The off diagonal components of \bbeta, are varied in a similar
fashion to \br. However new connections are added, or old ones
removed according to $\lfloor 1/p\rfloor$, where $p\in(-2,2)$ is
chosen from a uniform distribution. The values on the new connections
are chosen from the same initial distribution that the off diagonal
values where originally set with, ie the range
$\left.\min_{i\neq j}\beta_{ij}\right|_{t=0}$ to
$\left.\max_{i\neq j}\beta_{ij}\right|_{t=0}$. Since condition
(\ref{boundedness3}) is computationally expensive, we use a slightly stronger
criterion that is sufficient, computationally tractable yet still
allows ``interesting'' non-definite matrix behaviour namely that the sum
$\beta_{ij}+\beta_{ji}$ should be non positive.

\item \bmu\ must be positive, so should evolve according to the
log-normal distribution like the diagonal components of
\bbeta. Similar to \bbeta, it is a catastrophe to allow \bmu\ to
become arbitrarily large. In the real world, mutation normally exists
at some fixed background rate --- species can reduce the level of
mutation by improving their genetic repair algorithms. In ecolab, this
ceiling on \bmu\ is given by the \verb|mutation(random,maxval)| variable.

\end{enumerate}

\subsection{Typical Results}

\begin{figure} 
\epsfbox{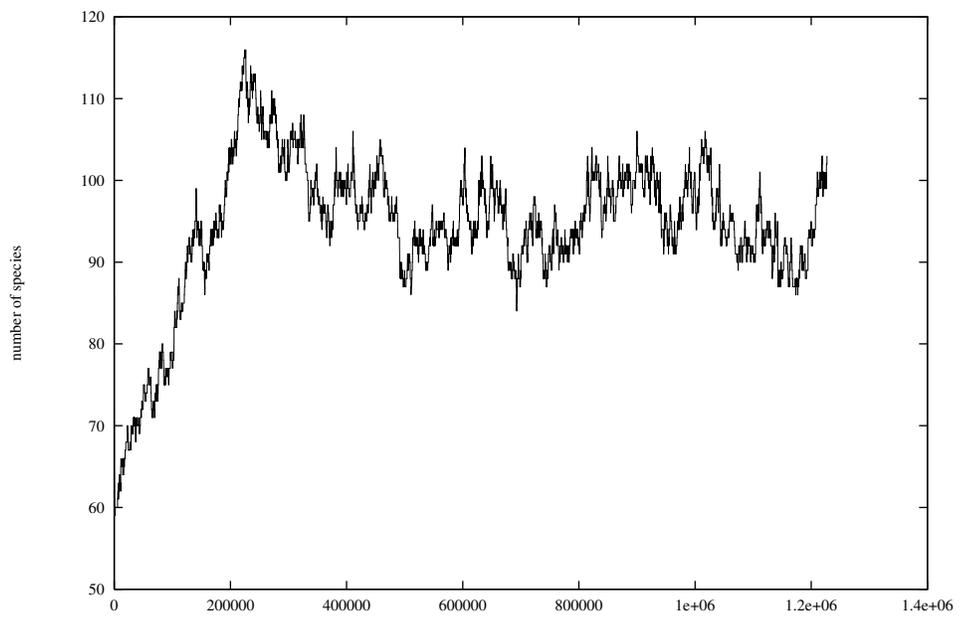}
\caption{\nsp\ as a function of time step}
\label{nsp}
\end{figure}

Figure \ref{nsp} shows the time behaviour for the number of species in
the ecosystem for a typical run. The phenotypic parameters were seeded
randomly in the ranges $-0.005\leq\br\leq0.01$,
$-5\times10^{-5}\leq\bbeta_{\rm diag}\leq-1\times10^{-4}$,
$-0.001\leq\bbeta_{\rm offdiag}\leq0.001$ and $0\leq\bmu\leq0.09$. The
\br\ and \bbeta\ values were chosen so that several hundred
individuals will be supported in the case of a single species system,
and the offdiagonal terms large enough to permit interesting
interactions between species, but not so large that the system
collapsed to zero immediately.  $\rho$ was set at $10^4$, which was
chosen by examining the histogram of differences between all the
species. If $\rho$ was too small, then a species' mutant offspring
would be too similar to its parent to be really a new species. This
shows up as a peak at small separation values of the histogram, which
shouldn't be there according to the law of competitive exclusion.

The system rapidly evolves to one of the fixed points (by a massive
extinction event!) with a negative definite \bbeta. Over time,
mutations build up in the system, decreasing the stability of the
system. What then follows are periods of episodic extinctions, and
system growth through speciation. This is an example of {\em self
  organised criticality}\cite{Bak-etal88}, and gives rise to power law
behaviour.

\begin{figure}
\epsfbox{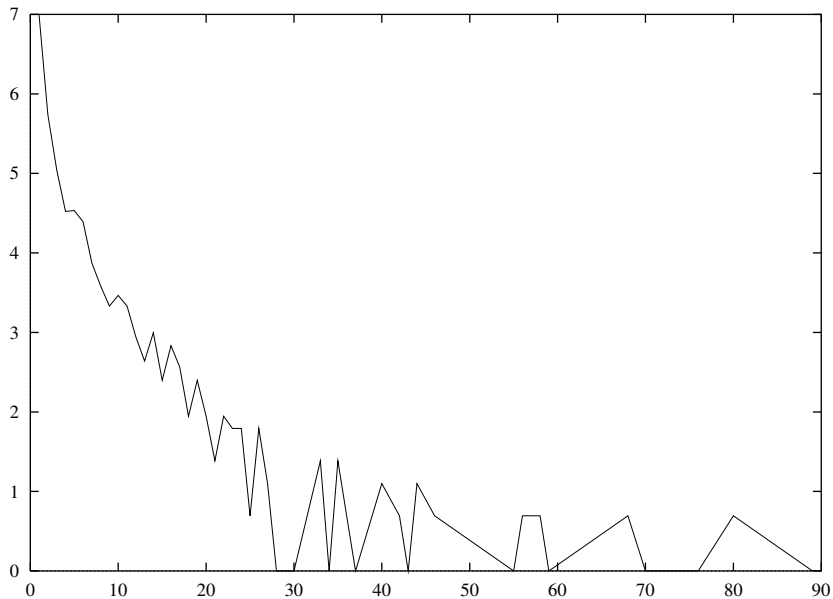}
\caption{Distribution of species lifetimes on a log-linear
  plot. Distribution is unnormalised. Horizontal scale is the
  natural logarithm of species lifetime in timesteps.}
\label{log-lin}
\end{figure}

\begin{figure}
\epsfbox{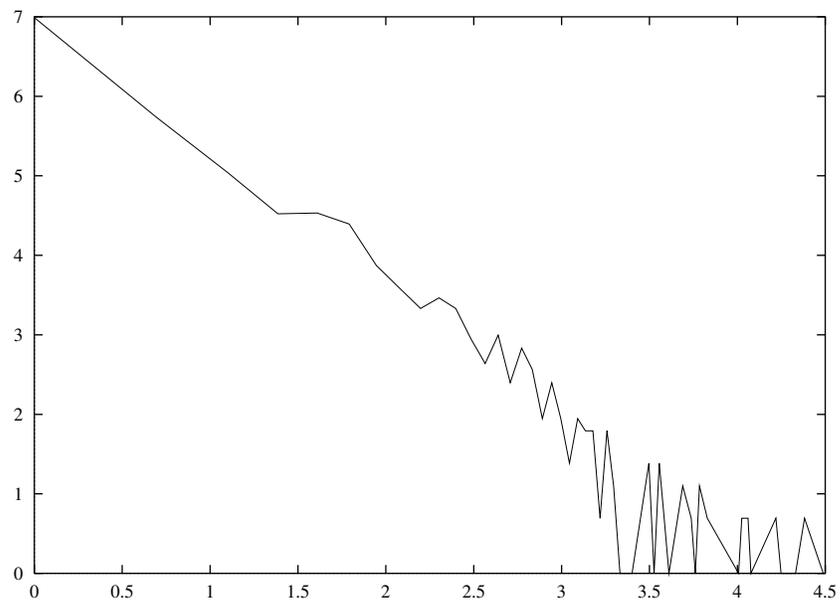}
\caption{Distribution of species lifetimes on a log-log
  plot. Vertical scale is the natural logarithm of that in figure
  \ref{log-lin}. Horizontal scale is the same.}
\label{log-log}
\end{figure}

Do we see the same power law behaviour observed by
others?\cite{Bak-Sneppen93,Adami-Brown94} The answer is emphatically
yes. If speciation and extinction events occurred uniformly throughout
history, as one might na\"\i{}vely expect, one would expect a Poisson
distribution for species lifetimes. On a log-linear plot, this would
be a straight line. Alternatively, if a power law spectrum was
evident, the log-log plot would be straight. The two plots are shown
in figures \ref{log-lin} and \ref{log-log}. Effectively, this is
telling us that not only is there not a stable ecological equilibrium,
there isn't even a steady state, whereby extinctions are balanced by
speciation (a common ecological assumption). 

This model then is a concrete example of the self-organised
criticality predicted in these types of systems in
section\ref{limits}. The next section examines a possible economic
model that is analogous to Ecolab, and could even be implemented using
the same simulation software. It would be surprising if the dynamics
weren't critically self-organised.

\section{Building an Economic Dynamics}

Many inferential similarities can be drawn between the biological
evolutionary model of Ecolab and the processes of a capitalist
economy. The obvious analogy for a biological species is a product,
and for Darwinian evolution the process of technological change. I
consider a model economics ({\em Econolab}) based on the insights of
von Neumann, one of the founders of complexity theory, who introduced
{\em von Neumann Technology} in the late
1930s\cite{vonNeumann37,vonNeumann45}. In this model economy, there is
a set of {\em commodities} labeled $i\in{\cal N}=\{1\ldots N\}$), and
a set of {\em technologies} or {\em processes} labeled $m\in{\cal
  M}=\{1\ldots M\}$. Each process has an activity $z_m$, input
coefficients $a_{mi}$ and output coefficients $b_{mi}$, such that in
one time step, $a_{mi}z_m$ of commodity $i$ (amongst others) is
consumed to produce $b_{mj}z_j$ of commodity $j$ (amongst others). The
coefficients $a_{mi}$ and $b_{mi}$ may be zero for some values of $m$
and $i$, corresponding respectively to processes that do not require a
particular input, or do not produce a particular output. This differs
from von Neumann's original approach, and is more in line with that of
Kemeny, Morgenstern and Thompson\cite{Kemeny-etal56}.
Blatt\cite{Blatt83} gives a good introduction to this model,
discussing it flexibility in dealing with a range of economic
processes. In the words of Blatt (p67):
\begin{quote}
The von Neumann work is a great achievement of mathematical model
building in dynamic economics. It is the best available theory of
capital and of rate of return.
\end{quote}

That said, there are many issues of significance in capitalism which
are not captured in the von Neumann method, and which cannot be
modeled in an initial rendition of Econolab. These include effective
demand\cite{Chiarella98}, income distribution, variable
capacity and utilisation, credit and debt\cite{Keen98,Andresson98}.

To relate von Neumann's work back to the Ecolab ecological model, the
input/output coefficients $a_{mi}$/$b_{mi}$ are fixed like the $r_i$,
$\beta_{ij}$ or equation (\ref{lotka-volterra}), and $z_m$ is a free
variable like $n_i$. In von Neumann's work, the dynamics is imposed in
the form of an exponential growth condition:
\begin{equation}\label{exp-growth}
z_m(t+1)=\alpha z_m(t) \;\; \forall m\in{\cal M}
\end{equation}

However, rather than assuming a particular form for the dynamics, we should be
looking for a first order differential equation (or its difference
equation equivalent) that describes the dynamics. Consider the
monetary value of capital
$K_m$ associated with process $m$. The rate of change of this capital
may be written:
\begin{equation}\label{cap-rate-change}
\dot{K}_m=z_m\left(\sum_{i=1}^N b_{mi}p_i - \sum_{i=1}^N a_{mi}p_i\right),
\end{equation}
where $p_i$ is the price of
commodity $i$. This has introduced two new sets of free variables
$K_m$ and $p_i$, for which we need to find closure relations. Clearly,
activity is limited by the availability of capital (we do not allow
the possibility of credit here):
\begin{equation}
\sum_{i=1}^N a_{mi}p_iz_m \leq K_m
\end{equation}

For simplicity, let us assume that each process invests a fixed
proportion of its capital into production, i.e.
\begin{equation}\label{fixed-capital}
\sum_{i=1}^N a_{mi}p_iz_m = \kappa_m K_m, \;\;  \exists\kappa_m: 0<\kappa_m\leq1
\end{equation}
Substituting (\ref{fixed-capital}) into (\ref{cap-rate-change}) gives
\begin{equation}\label{activity-rate-change}
\dot z_m=\kappa_m\left(
\frac
   {\sum_{i=1}^N b_{mi}p_i}
   {\sum_{i=1}^N a_{mi}p_i}
-1
\right) z_m.
\end{equation}
This then, is a model dynamics analogous to the Lotka-Volterra
equation (\ref{lotka-volterra}).  If price is a fixed quantity (as
assumed in von Neumann theory) then (\ref{activity-rate-change}) is
equivalent to the {\em ansatz} (\ref{exp-growth}). This is the
equilibrium situation, rather like assuming that $\bn=\hat{\bn}$. 

In reality, prices are not fixed, and must have their own dynamics.
The simplest way to do this is to look for a closure relation, that
relates prices to activities. The neoclassical and Austrian traditions
propose that price dynamics should act as a negative feedback on the
activity dynamics (eq. (\ref{activity-rate-change})), whereas the P-K
and Sraffian tradition do not see price as an equilibrating mechanism
\cite{Sraffa26}. In this work, however, we propose an ansatz on the
form of the negative feedback, in a similar fashion to the ansatz used
by Nos\'e and Hoover\cite{Hoover85} to describe the thermostat that
regulates the temperature of a non-equilibrium steady state system in
a heat bath:
\begin{equation}\label{price-dynamics}
\dot p_i = \pi_i\left(
\frac{\rm demand}{\rm supply}-1
\right) p_i =
\pi_i\left(
\frac
  {\sum_{m=1}^M a_{mi}z_m}
  {\sum_{m=1}^M b_{mi}z_m}
-1
\right)p_i.
\end{equation}
This differs from von Neumann, who assumes that demand never exceeds supply, and if supply
exceeds demand (i.e. a surplus), then the commodity is free
($p_i=0$). This would imply $\dot p_i=0$, freezing prices. In effect this
makes the system very stiff --- equation (\ref{price-dynamics}) softens
the dynamics with $\pi_i$ controlling the stiffness.

\section{Adding Evolution}

Now that we have an economic dynamics established, we need to consider
how to develop an analogy between ecological and economic evolution.
By direct analogy with Ecolab, it is clear that when a process
exhausts its capital ($K_m=0$), it forever remains that way, so this
is equivalent to extinction in ecosystems. Adding new processes and
commodities is conceptually easy. Blatt p57--58\cite{Blatt83}:
\begin{quote}
What about technological progress? This can be included by assuming
that the list of activities $m=1, 2, \ldots, M$ is not final, but new
activities may be invented and hence become available for use, as time
goes on. This makes the total number of processes a function of time:
$M=M(t)$\ldots Von Neumann himself developed his theory on the basis of an
unchanged technology (all input coefficients, output coefficients
and the number of processes $M$ are constant in time), and his
successors have done the same. The inclusion of technological progress
appears to us to be a highly interesting avenue for further
exploration.
\end{quote}

The difficulty is deciding how to choose new coefficients $a_{mi},
b_{mi}, \kappa_i$ and $\pi_i$ when a new process is added. There is no
genotype of a process --- the closest thing to it is Dawkins's
meme, and there is no genetic algorithm theory of the
meme. Clearly new processes arise evolutionarily, with the new
processes modeled on the old. The new coefficients will be varied
randomly about the old values according to some kind of central
distribution.

Recent results from Ecolab indicate that the emergent dynamics of the
system is rather insensitive to the specific type of mutation
algorithm chosen. Work is currently under way to classify exactly what
effects different assumptions make.

In 1962, Arrow\cite{Arrow62} pointed out that the cost per unit for
production of an artifact falls as an inverse power of the number of
units produced: 
\begin{displaymath}
{\rm cost}/{\rm unit} \propto N^{-a}
\end{displaymath}
This power law is most likely a consequence of the dynamics of technological
innovation, relating to the statistical
properties of the underlying ``fitness'' landscape, as it can be seen
in Kauffman $NK$ model\cite{Kauffman93}. Presumably an evolutionary
algorithm that searches process (and commodity) space according to the
same power law would be optimally matched to generating
change, however another search algorithm would probably generate the
same distribution of successful innovations, albeit on a different
temporal scale. It should also be pointed out that large changes of
process are likely to cost proportionally more than smaller changes.
As any research budget is finite, the distribution of process
improvements must therefore be finitely integrable (have a finite area
underneath the curve), which the power law distribution is not, but
the normal (Gaussian) distribution is.

\section{Conclusion}

Economics is clearly a dynamic process, which given its complexity will
be poorly described by a linear approximation about a stable
equilibrium. Rather the properties of the equilibrium will be
determined by cultural evolution which operates over a longer
timescale than economics. It is likely that cultural evolution will
produce a self-organised critical system, and this would be one of the
first questions to study. Other questions that might be looked at
include looking for evidence for the Arrow law, and looking for
analogues to various biological laws, such as the species-area law
\footnote{the number of species on an island is related by a power law to
the area of the island}, and dependence of biodiversity with latitude.

Perhaps the most important point I would like to make is that rather
than studying a finite dimensional dynamical system, we should be
studying what might be called ``open dimensional'' dynamical systems,
where the number of degrees of freedom is finite, but not fixed at any
point in time. These systems must lie between finite dimensional
spaces and infinite dimensional ``functional analysis'' type
spaces. Only then might we achieve Marshall's economic biology, and
have an understanding of why economic systems have evolved to be the way
they are.

\bibliographystyle{plain}
\bibliography{rus}

\begin{thebibliography}{10}

\bibitem{Adami-Brown94}
Chris Adami and C.~Titus Brown.
\newblock Evolutionary learning in the 2d artificial life system ``avida''.
\newblock In R.~Brooks and P.~Maes, editors, {\em Artificial Life IV}. MIT
  Press, 1994.

\bibitem{Arrow62}
K.~Arrow.
\newblock The economic implications of learning by doing.
\newblock {\em Review of Economic Studies}, 29:166, 1962.

\bibitem{Bak-etal88}
P.~Bak, C.~Tang, and K.~Wiesenfeld.
\newblock Self-organised criticality.
\newblock {\em Phys. Rev. A}, 38:364, 1988.

\bibitem{Bak-Sneppen93}
Per bak and Kim Sneppen.
\newblock Puntuated equlibrium and criticality in a simple model of evolution.
\newblock {\em Phys. Rev. Lett.}, 71:4083, 1993.

\bibitem{Blatt83}
J.~M. Blatt.
\newblock {\em Dynamic Economic Systems: A Post-Keynesian Approach}.
\newblock M. E. Sharpe, New york, 1983.

\bibitem{Bruckner-etal93}
E.~Bruckner, W.~Ebeling, J.~Montano, and A.~Scharnhorst.
\newblock Technological innovations --- a self-organisation approach.
\newblock Technical Report FS II 93--302, Wissenschaftzentrum Berlin fuer
  Sozialforschung, Reichpietschufer 50, 1000 Berlin 30, 1993.

\bibitem{Case91}
T.~J. Case.
\newblock Invasion resistance, species build-up and community collapse in
  metapopulation models with interspecies competition.
\newblock {\em Bio. J. Linnean Soc.}, 42:239--266, 1991.

\bibitem{Hirsch-Smale74}
M.~W. Hirsch and S.~Smale.
\newblock {\em Differential equations, dynamical systems, and linear algebra}.
\newblock Academic Press, New York, 1974.

\bibitem{Hofbauer-etal87}
J.~Hofbauer, V.~Hutson, and W.~Jansen.
\newblock Coexistence for systems governed by difference equations of
  lotka-volterra type.
\newblock {\em J. Math, Biol.}, 25:553--570, 1987.

\bibitem{Hoover85}
W.~G. Hoover.
\newblock {\em Phys. Rev. A}, 31:1695, 1985.

\bibitem{Jansen87}
W.~Jansen.
\newblock A permanence theorem for replicator and lotka-volterra systems.
\newblock {\em J. Math. Biol.}, 25:411--422, 1987.

\bibitem{Kauffman-Macready95}
S.~Kauffman and W.~Macready.
\newblock Technological evolution and adapative organizations.
\newblock {\em Complexity}, 1:26, 1995.

\bibitem{Kauffman93}
Stuart~A. Kauffman.
\newblock {\em The Origins of Order: Self Organization and Selection in
  Evolution}.
\newblock Oxford UP, 1993.

\bibitem{Kemeny-etal56}
J.~G. Kemeny, O.~Morgenstern, and G.~L. Thompson.
\newblock A generalization of the von neumann model of an expanding economy.
\newblock {\em Econometrica}, 24:115--135, 1956.

\bibitem{Law-Blackford92}
Richard Law and Jerry~C. Blackford.
\newblock Self-assembling food webs: A global viewpoint of coexistence of
  species in lotka-volterra communitites.
\newblock {\em Ecology}, 73:567--578, 1992.

\bibitem{Marshall20}
Alfred Marshall.
\newblock {\em Principles of Economics}.
\newblock Macmillan, London, 8th edition, 1920.

\bibitem{May76b}
R.~M. May.
\newblock Simple mathematical model with very complicated dynamics.
\newblock {\em Nature}, 26:457, 1976.

\bibitem{May74}
Robert~M. May.
\newblock {\em Stability and Complexity in Model Ecosystems}.
\newblock Princeton University Press, Princeton, New Jersey, 1974.

\bibitem{Maynard-Smith74}
J.~Maynard~Smith.
\newblock {\em Models in Ecology}.
\newblock Cambridge University Press, London, 1974.

\bibitem{Muller89}
Richard Muller.
\newblock {\em Nemesis: The Death Star}.
\newblock Heinemann, London, 1989.

\bibitem{Ousterhout94}
J.~K. Ousterhout.
\newblock {\em Tcl and the Tk Toolkit}.
\newblock Addison-Wesley, 1994.

\bibitem{Raup86}
David~M. Raup.
\newblock {\em The Nemesis Affair}.
\newblock Norton, New York, 1986.

\bibitem{Raup91}
David~M. Raup.
\newblock {\em Extinction: Bad Genes or Bad Luck?}
\newblock Norton, New York, 1991.

\bibitem{Raup-Sepkowski86}
David~M. Raup and J.~John Sepkowski, Jr.
\newblock Periodic extinctions of families and genera.
\newblock {\em Science}, 231:833--836, 1986.

\bibitem{Schroeder91}
M.~Schroeder.
\newblock {\em Fractals, Chaos, Power Laws}.
\newblock Freeman, New York, 1991.

\bibitem{Sepkowski90}
J.~John Sepkowski, Jr.
\newblock The taxonomic structure of periodic extinction.
\newblock In Virgil~L. Sharpton and Peter~D. Ward, editors, {\em Global
  Catastrophes in Earth History}, number 247 in The Geological Society of
  America Special Papers, pages 33--44. The Geological Society of America,
  Boulder, Colorado, 1990.

\bibitem{Standish96}
R.~K. Standish.
\newblock Ecolab: Where to now?
\newblock In R.~Stocker, H.~Jelinek, B.~Durnota, and T.~Bossomeier, editors,
  {\em Complex Systems: From Local Interaction to Global Phenomena}. IOS, 1996.
\newblock also Proceedings of AlifeV, Nara Japan, May 1996 and {\em Complexity
  International}, vol. 3, http://www.csu.edu.au/ci.

\bibitem{Ecolab-Tech-Report}
Russell~K. Standish.
\newblock Ecolab documentation.
\newblock Available at http://parallel.acsu.unsw.edu.au/rks/ecolab.html.

\bibitem{Standish94}
Russell~K. Standish.
\newblock Population models with random embryologies as a paradigm for
  evolution.
\newblock In Russel~J. Stonier and Xing~Huo Yu, editors, {\em Complex Systems:
  Mechanism of Adaption}. IOS Press, Amsterdam, 1994.
\newblock also {\em Complexity International}, vol. 2,
  http://www.csu.edu.au/ci.

\bibitem{Strobeck73}
C.~Strobeck.
\newblock $n$ species competition.
\newblock {\em Ecology}, 54:650--654, 1973.

\bibitem{Thompson92}
C.~J. Thompson.
\newblock Chaos in economics and management.
\newblock In R.~L. Dewar and B.~I Henry, editors, {\em Nonlinear Dynamics and
  Chaos}, pages 213--229. World Scientific, Singapore, 1992.

\bibitem{vonNeumann37}
J.~von Neumann.
\newblock A model of general economic equilibrium and a generalization of
  brouwer's fixed point theorem.
\newblock In K.~Menger, editor, {\em Ergebnisse Eines Mathematischen
  Kolloquiums}. Vienna, 1937.

\bibitem{vonNeumann45}
J.~von Neumann.
\newblock A model of general economic equilibrium.
\newblock {\em Review of Economic Studies}, 13:1--9, 1945.

\end{thebibliography}

\end{document}